\documentclass[aps,pre,preprint,groupedaddress,superscriptaddress,showpacs,a4paper,floatfix]{revtex4-1}

\usepackage{amsmath,amssymb,graphicx,graphics,bm}
\usepackage{hyperref}
\usepackage{graphics,graphicx}
\usepackage{amsmath,amssymb}

\begin{document}

\title{Long-tailed dissipationless hydromechanics: weak thermalization
and ergodicity breaking}

\author{Giuseppe Procopio$^1$,  Chiara Pezzotti$^1$ and Massimiliano Giona}
\email[corresponding author:]{massimiliano.giona@uniroma1.it}
\affiliation{Dipartimento di Ingegneria Chimica, Materiali, Ambiente La Sapienza Universit\`a di Roma\\ Via Eudossiana 18, 00184 Roma, Italy}

\date{\today}

\begin{abstract}
We analyze the dynamic properties of dissipationless Generalized
Langevin Equations in  the presence of fluid inertial kernels
possesing power-law tails, $k(t) \sim t^{-\kappa}$. While for
$\kappa >1$ the dynamics is manifestly non ergodic, no thermalization
occurs, and particle motion is ballistic, new phenomena arise for
$0 < \kappa <1$. In this case, a form of weak thermalization appears 
in the presence of thermal/hydrodynamic fluctuations and attractive
potentials. However, the absence of dissipation clearly emerges once 
an external constant force is applied: an asymptotic settling velocity
cannot  be achieved
as the expected value of the 
particle velocity diverges.
\end{abstract}

\maketitle

\noindent
\section{Introduction}  
\label{sec1}
Starting from the work by Kubo \cite{kubo1,kubo2}, fluctuation-dissipation
theory has been extended to  fluctuating dynamics possessing
memory effects, and the theory of Generalized
Langevin Equations (GLE) has progressively acquired an ever more prominent role
in statistical physics \cite{gle1,gle2,gle3,gle4}.
Amongst  the rich phenomenology characterizing the
dynamics of GLEs \cite{gen}, the occurrence of ergodicity
breaking and its characterization has attracted 
a considerable attention \cite{br1,br2,br3,br4,plyukhin}. This is partially due
to the fact that ergodicity breaking in GLE occurs under 
slightly different  conditions with respect to the kinematic
counterpart, e.g. continuous random walk models \cite{ct1,ct2,ct3}.
In the latter case, this is mainly due to long-range correlations
and diverging transition time.
In the case of GLE, two main mechanisms have been
so far identified: the lack of dissipation \cite{br1,br2,br3,br4}, and the
the break-up of dissipative stability in the meaning addressed in \cite{gpp1}.
The second case occurs even  in the presence of a non vanishing
positive and bounded friction factor  \cite{gpp2} and finds
in the Plyukhin model  \cite{plyukhin} a simple and beautiful example.

Although  GLEs arise in a variety of physical system, particle
hydromechanics in a fluid medium, in  the operating conditions at which
thermal/hydrodynamic fluctuations are relevant (micrometric particles),
is probably the paradigmatic physical problem in which GLE theory applies
\cite{kubo2,gen,gpp1,hb1}.
Consequently, it is not surprising that particle hydromechanics
can be used as
the backbone of a theory of memory effects in GLE, discriminating
and dissecting their different physical origins.

In the case of particle motion in macroscopically still fluids
(Brownian motion), memory effects arise from two different
physical phenomena: viscoelasticity and fluid inertia.
Viscoelasticity is associated with dissipative memory effects
in the constitutive equations  relating shear-stress  to deformation
tensors \cite{rheol1,rheol2}. The second mechanism involves fluid inertial
effects \cite{landau} deriving from the description of medium fluid hydrodynamics
via the time dependent Stokes equation, made necessary by
the high value of the Strouhal number deriving by the high frequency
of oscillations characterizing thermal fluctuations \cite{pofgiona}. In the
Newtonian and incompressible case,  memory fluid inertial
effects appear in the form of the  memory Basset force (characterized
by a $1/\sqrt{t}$-memory kernel) \cite{landau}.

In this article we analyzing the dynamic properties of 
dissipationless GLEs characterized  by fluid inertial kernels
(see further Section \ref{sec2}) possessing long-lasting
power-law tails
\begin{equation}
k(t) \sim t^{-\kappa}
\label{eq1}
\end{equation}
for large $t$, as new, and partially unexpected features
arise for small values of the exponent $\kappa <1$.

For $\kappa>1$, ergodicity is broken, the system does
not display thermatization as regards momentum transfer, 
and particle motion is,  for most of the initial conditions,
purely ballistic.  Conversely, for $\kappa < 1$ some form
of weak thermalization (defined in Section \ref{sec4}) 
occurs, and this is associated
with a superdiffusive motion.
Moreover, the change from ergodic to non ergodic behavior
near the critical point $\kappa_c=1$
corresponds to a phase transition, or equivalently
to a bifurcation in the meaning of dynamical system
theory associated with the exchange of stability 
(see Section \ref{sec3}).

In point of fact, the occurrence of some form of thermalization
for $\kappa<1$ motivates  a finer
characterization of the behavior
of physical systems in the presence of
different forces (stochastic thermal forces, 
conservative forces stemming from a potential, bounded non conservative
forces, etc.), and a deeper analysis on the role
of dissipation in irreversible processes and on the singular
limits characterizing the dissipationless case, as discussed
in Section \ref{sec4}.

The article is organized as follows.
Section \ref{sec2} defines  the mathematical and hydromechanic
setting of the problem and its modal formulation
consistent with Kubo fluctuation-dissipation theory.
Section \ref{sec3} develops the novel part, associated
with the qualitative dynamic behavior (ergodicity and its breaking)
 depending on the exponent $\kappa$ entering
eq. (\ref{eq1}),
 addressing in detail the characterization of  the phase transition/bifurcation
occurring at $\kappa=1$. Section \ref{sec4} addresses
further the role of dissipation and irreversibility
in the presence of thermal fluctuations, external forces and
potentials.  The difference between weak thermalization and strong
dissipative behavior is carefully discussed.

\section{Setting of the problem}
\label{sec2}

Consider the hydromechanics of a  spherical particle of mass $m$
in a still fluid (sphericity essentially implies the isotropy
of fluid-particle interactions). The equation of motion
for the particle velocity $v(t)$ reads  as \cite{pofgiona}
\begin{equation}
m \, \frac{d v(t)}{d t} = - \int_0^\infty h(t-\tau) \, v(\tau) \, d \tau
- \int_0^t k(t-\tau) \, \left ( \frac{d v(\tau)}{d \tau} + v(0) \, \delta(\tau)
\right )\, d \tau + R(t)
\label{eq2}
\end{equation}
where $R(t)$ is the thermal/hydrodynamic fluctuational force,
defined by the Kubo fluctuation-dissipation relations \cite{kubo1,kubo2},
which essentially means the fulfillment of the Langevin condition
$\langle R(t) \, v(\tau) \rangle_{\rm eq}=0$ for $\tau\leq t$,
where $\langle \cdot \rangle_{\rm eq}$ indicates the
expected value with respect to the equilibrium probability measure.

The   GLE eq. (\ref{eq2}) is characterized by two kernels:
the dissipative kernel $h(t)$, and the fluid-inertial
kernel $k(t)$ \cite{pofgiona}.    
The dissipative kernel involved in fluid-particle
interactions is non negative \cite{gpp2}, i.e. $h(t) \geq 0$ (for other
physical phenomelogies this condition can be relaxed \cite{plyukhin}),
and this stems from rheology \cite{rheol1,rheol2}. Its integral
$\eta_\infty=\int_0^\infty h(t) \, d t$, if bounded and different
from zero,  enters the global Stokes-Einstein relation for
the long-term particle diffusivity $D$ at constant
temperature $T$, $D \, \eta_\infty=k_B \, T$, where
$k_B$ is the Boltzmann constant. Conversely, the fluid inertial
term, defined by the kernel $k(t)$ does not contribute to dissipation. It
modulates the shape of the velocity autocorrelation function \cite{raizen1,raizen2}.
From hydrodynamic analysis, i.e. for the few cases  in which $k(t)$
is analytically available (Newtonian fluids, Maxwell fluids) \cite{landau,peppevisco}, it can be
assumed that $k(t) \geq 0$, albeit there is no conclusive result
in this regard.

Henceforth, let ut consider eq. (\ref{eq2}) in dimensionless form,
by rescaling time and velocity so that $\langle v^2 \rangle_{\rm eq}=1$
would correspond to the Maxwellian equipartition relation, and
$h(t)=0$ i.e. no dissipation.
If  one further assume $k(t=0)$ bounded, as for any viscoelastic
fluids \cite{peppevisco}, the fluid-inertial  term can be
reworked by parts, so that the dimensioless version of  eq. (\ref{eq2})
(in dimensionless form $m=1$, and we keep the notation $t$, $v(t)$, $R(t)$ and $k(t)$
to indicate the dimensionless  time, velocity, random force
 and inertial kernel), takes the
form.
\begin{equation}
\frac{d v(t)}{d t}= - k(0) \, v(t) + \int_0^t \frac{d k(t-\tau)}{d \tau}
\, v(\tau) \,d \tau + R(t)
\label{eq3}
\end{equation}

We are interested in the case where $k(t)$ possesses long-term power-law
behavior. To this end, once can consider the class of kernels
 defined by the analytic function,
\begin{equation}
g_\kappa(t)=\sum_{n=1}^\infty \frac{1}{ 2^{n \, \kappa}} \, e^{-t/2^n}
\sim t^{-\kappa}
\label{eq4} 
\end{equation}
with $\kappa>0$, i.e. $k(t)=g_\kappa(t)$.  The truncated representations
\begin{equation}
g_\kappa(t;N)= \sum_{n=1}^N \frac{1}{ 2^{n \, \kappa}} \, e^{-t/2^n}
\label{eq5}
\end{equation}
for $N$ large enough, but still order of $10^1$, provide accurate
power-law scalings over several decades.
\begin{figure}
\includegraphics[width=10cm]{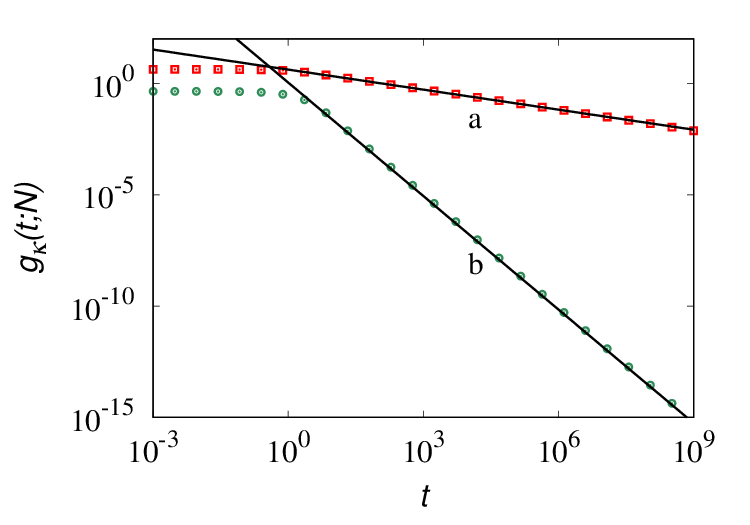}
\caption{Functions $g_\kappa(t;N)$ vs $t$ defined by eq. (\ref{eq5})
with $N=40$. Symbols ($\square$) refer to $\kappa=0.3$,
($\circ$) to $\kappa=1.7$. Solids lines (a) and (b)
represent the scalings $g_\kappa(t;N) \sim t^{-\kappa}$ for
the two values of the exponent $\kappa$ considered.}
 \label{Fig1}
\end{figure}

Specifically, figure \ref{Fig1} depicts the behavior of
$g_\kappa(t;N)$ for $N=40$ at two different values of $\kappa$.
It can be observed that the power-law scaling holds up to $t=10^9$,
and this value is the upper integration limit  in the
simulations of the dynamics reported in the remainder (see Section \ref{sec3}).
Henceforth we assume $k(t)=g_\kappa(t;N)$ with $N$
large enough to ensure the power-law scaling of the kernel
over the timescales involved.
Set
\begin{equation}
\gamma_n = \frac{1}{2^{n \, \kappa}} \, , \quad
\mu_n= \frac{1}{2^n}\, , \quad \Gamma= \sum_{n=1}^N \gamma_n
\label{eq6}
\end{equation}
$n=1,\dots,N$, so that $k(t)=\sum_{n=1}^N \gamma_n e^{-\mu_n \, t}$. Eq. (\ref{eq3}) with the kernel $k(t)$ given
by eq. (\ref{eq4}) can be  recast in the modal expansion (Markovian embedding)
\cite{gpp1}
involving the auxiliary variables $z_n(t)$, $n=1,\dots,N$,
\begin{eqnarray}
\frac{d v(t)}{d t} & = & - \Gamma \, v(t) + \sum_{n=1}^N \gamma_n \, \mu_n \,
z_n(t) + \sqrt{2} \, \sum_{n=1}^N d_n \, \xi_n(t) \nonumber \\
\frac{d z_n(t)}{d t} & = & - \mu_n \, z_n(t) + v(t) +
\sqrt{2} \, c_n \, \xi_n(t) \, , \;\;\;n=1,\dots,N
\label{eq7}
\end{eqnarray}
where $\xi_n(t)=d w_n(t)/dt$, $n=1,\dots,N$ are the distributional derivatives
of  independent Wiener processes $w_n(t)$, so that
$\langle \xi_m(t) \xi_n(t^\prime) \rangle = \delta_{mn} \, \delta(t-t^\prime)$,
and 
\begin{equation}
d_n= \sqrt{\gamma_n} \, , \quad c_n= - \frac{1}{\sqrt{\gamma_n}} \, ,
\quad n=1,\dots,N
\label{eq8}
\end{equation}
Eq. (\ref{eq6}) encompasses also the representation of the thermal force $R(t)$
consistently with the Kubo fluctuation-dissipation theory, and  the
expression for the coefficients $c_n$ and $d_n$ modulating
the intensity of the Wiener stochastic forcings has been
derived from the Langevin condition (see Appendix A).

\section{Main results}
\label{sec3}
This Section reports the scaling theory and the main qualitative
properties of the GLE dynamics eqs. (\ref{eq3}) or (\ref{eq7}).
To begin with, for $\kappa>1$ the dynamics is manifestly
non ergodic, as the statistical properties of
the particle velocity depend on the initial conditions.
This can be easily checked (see below and Appendix B)
by considering different initial conditions as
regards the   values of $v(t)$ and $z_n(t)$, $n=1,\dots,N$
at $t=0$, and observing that in general 
\begin{equation}
\lim_{t \rightarrow \infty} \langle  v^2(t) \rangle_0 \neq
\langle v^2 \rangle_{\rm eq}=1
\label{eq9}
\end{equation}
where $\langle \cdot \rangle_0$ refers to
the expected value referred to an initial
joint density $p_0(v_0,{\bf z}_0)$,
where ${\bf z}_0=(z_{1,0},\dots,z_{N,0})$.
Henceforth we will indicate with $\langle v^2(t) \rangle$ or 
$\langle v^2(t) \rangle_0$ the
square velocity variance as a function of time, starting from a given
initial statistical setting specified case by case.
And this should not be confused with $\langle v^2 \rangle_{\rm eq}$
that is the equilibrium value equal to $k_N T/m$ in the dimensional
formulation and to $1$ in the present nondimensional setting.

In a similar way, the velocity autocorrelation function
$C_{vv}^0(t)$, starting from specific initial conditions
defined by  the density $p_0(v_0,{\bf z}_0)$, is given by 
\begin{equation}
C_{vv}^0(t)= \int d v \int v \, v_0 \, p(v,t \, | \, v_0, 0) \,
p_0(v_0) \, d v_0
\label{eq10}
\end{equation}
where $p(v,t \, | \, v_0, 0)$ is the conditional
probability density of having the velocity $v$ at time $t$ if
the initial velocity  at time $t=0$  is $v_0$, and
$p_0(v_0)= \int p_0(v_0,{\bf z}_0) d {\bf z}_0$ is
the  initial marginal  velocity density.
In the non ergodic case, $C_{vv}^0(t)$ may be different from the
Kubo velocity autocorrelation function at equilibrium, indicated
as $C_{vv}(t)$ (see further eq. (\ref{eq13}) for its definition).

Conversely, for $\kappa<1$ a regular relaxation towards $\langle v^2 
\rangle_{\rm eq}$ is observed, as well the collapse
of $C_{vv}^0(t)$ towards $C_{vv}(t)$ for any initial
conditions, suggesting the validity of the ergodic hypothesis.

\subsection{Analysis}
\label{sec3_1}

\begin{figure}
\includegraphics[width=10cm]{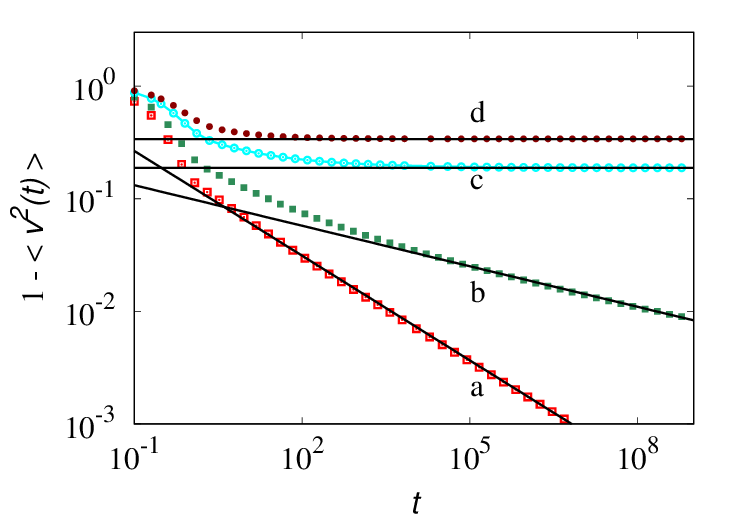}
\caption{$1- \langle v^2(t) \rangle$ vs $t$ for the dissipationless model
eq. (\ref{eq7}) obtained from the moment dynamics eq. (\ref{eq12}) (symbols)
starting from vanishing initial conditions $m_{vv}(0)=m_{vz_n}(0)=m_{z_nz_m}(0)=0$, $n,m=1,\dots,N$.  Symbols ($\square$ and line a) and ($\blacksquare$ and line b) refer to
$\kappa=0.7, \, 0.9$ respectively, while ($\circ$ and line c) and
($\bullet$ and line d) to $\kappa=1.3, \, 1.6$, respectively.}
 \label{Fig2}
\end{figure}

\begin{figure}
\includegraphics[width=16cm]{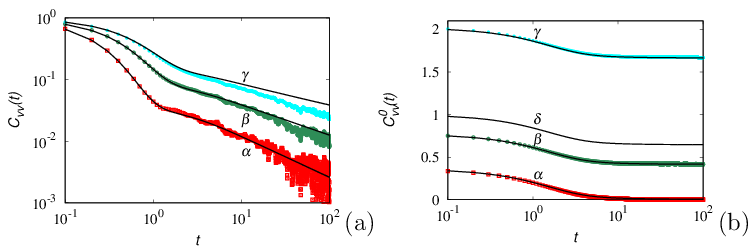}
\caption{Long-term velocity autocorrelation
function obtained from the averaging over an ensemble of 
realizations of eq. (\ref{eq7}). The initial
conditions are $z_n(0)=0$, $n=1,\dots,N$ while
$v(0)$ is a Gaussian random variable with  variance
$\sigma_0$. Panel (a)  reports data for
kernels with $\kappa<1$, $\sigma_0=0$. Symbols refer to
stochastic simulations with $\kappa=0.3, \, 0.5, \, 0.7$,
from line $\alpha$ to line $\gamma$, lines corresponds to
the Kubo velocity autocorrelation function.
Panel (b) refers to $\kappa=2.5$. Symbols and lines, from $\alpha$ to
$\gamma$, correspond to different values of $\sigma_0=0,\,1,\,2$.
Line ($\delta$) corresponds to the Kubo velocity autocorrelation
function. }
 \label{Fig3}
\end{figure}

Given the modal representation of the GLE eq. (\ref{eq7}),
the associated Fokker-Planck equation for the density $p(v,{\bf z},t)$,
where ${\bf z}=(z_1,\dots,z_N)$, reads
\begin{eqnarray}
\frac{\partial p}{\partial t} & = & \Gamma \frac{\partial (v \, p)}{\partial
v} - \sum_{n=1}^N \gamma_n \, \mu_n z_n \frac{\partial p}{\partial v}
+ \sum_{n=1}^N \mu_n \frac{\partial (z_n \, p)}{\partial z_n}
\nonumber \\
& - & \sum_{n=1}^N v \, \frac{\partial p}{\partial z_n}
+ \sum_{n=1}^N d_n^2  \, \frac{\partial^2 p}{\partial v^2}
+ 2 \sum_{n=1}^N c_n \, d _n  \, \frac{\partial^2 p}{\partial v \partial
z_n} + \sum_{n=1}^N c_n^2 \, \frac{\partial^2 p}{\partial z_n^2}
\label{eq11}
\end{eqnarray}
from which the evolution equations of the second-order 
moments follow
\begin{eqnarray}
\frac{d m_{vv}(t)}{d t} &= & - 2 \, \Gamma \, m_{vv}(t)+ 2 \, \sum_{n=1}^N
\gamma_n \, \mu_n \, m_{vz_n}(t) +  2 \sum_{n=1}^N d_n^2 \nonumber \\
\frac{d m_{v z_m}(t)}{d t} &= &  - \Gamma \, m_{v z_m}(t)+ \sum_{n=1}^N
\gamma_n \, \mu_n \,  m_{z_n z_m}(t) - \mu_m \, m_{v z_m}+ 2 \, c_n \, d_n
\label{eq12} \\
\frac{d m_{z_m z_p}(t)}{d t} &= & - (\mu_m+ \mu_p) \, m_{z_m z_p}(t)+ 
m_{v z_m}(t) + m_{v z_p}(t)+ 2 c_n^2 \, \delta_{mp}
\nonumber
\end{eqnarray}
The Kubo velocity autocorrelation function
follows directly from the modal expansion eq. (\ref{eq7}) and from the
Langevin condition, $\langle \xi(t) v(0) \rangle_{\rm eq}=0$,
\begin{eqnarray}
\frac{d C_{vv}(t)}{d t} & = & - \Gamma \, C_{vv}(t) +\sum_{n=1}^N
\gamma_n \, \mu_n \, C_{z_n v}(t) 
\label{eq13} \\
\frac{d C_{z_n v}(t)}{d t} & = & - \mu_n \, C_{z_n v}(t) + C_{vv}(t)
\nonumber
\end{eqnarray}
equipped with the initial conditions $C_{vv}(0)=\langle v^2 \rangle_{\rm eq}=1$,
$C_{z_n v}(0)=0$.
Figure \ref{Fig2} depicts the behavior of $\langle v^2(t) \rangle=m_{vv}(t)$,
starting from $m_{vv}(0)=m_{vz_n}(0)=m_{z_nz_m}(0)=0$,
$n,m=1,\dots,N$, for different values of the exponent $\kappa$. 
For $\kappa<1$ thermalization is achieved, and $\langle v^2(t) \rangle$
decay in a power-law way towards the equipartition result.
For $\kappa>1$ the long-term limit of $\langle v^2(t) \rangle$
converges towards a different (and smaller) value.
Analogous results in the presence of other initial
settings are reported in Appendix B.
A qualitatively similar result is obtained for the
velocity autocorrelation function, as depicted in figure \ref{Fig3}.
These data  are obtained from the stochastic simulation 
of eq. (\ref{eq6}) considering an ensemble of $N=10^5$ realizations
and  analyzing the correlation properties after a sufficiently
long transient.  

Below $\kappa=1$, the  velocity correlation properties
coincide with those predicted by fluctuation dissipation
theory while, for $\kappa>1$, these properties depend on
the initial conditions. In the latter case, we
observe for $C_{vv}^0(t)$ a translation of the
Kubo autocorrelation function, i.e. $C_{vv}^0(t)=C_{vv}(t)+K$
where the constant $K$ depends on the initial conditions.

We may conclude this analysis, with the further support of
the data reported in Appendix B, that, 
 for $\kappa<1$, eq. (\ref{eq7}) displays an ergodic
behavior and ergodicity breaks up at the critical point $\kappa_c=1$.

\subsection{Scaling analysis and anomalous diffusion}
\label{sec3_2}

A different diffusive behavior characterizes the system below
and above the critical value $\kappa_c$. To this end, let us
supplement eq. (\ref{eq7}) with the kinematic equation
\begin{equation}
\frac{d x(t)}{d t}= v(t)
\label{eq14}
\end{equation}
for the particle position $x(t)$.
The analysis of the diffusive anomalies can be viewed as the
extension of Mason's Generalized Stokes-Einstein relation
for dissipative viscoelastic fluids  \cite{mason1,mason2}
to the dissipationless
case.

Let $\widehat{C}_{vv}(s)$ be the Laplace transform of $C_{vv}(t)$,
where $s$ is the Laplace variable (henceforth we use the hatted notation
for Laplace transforms of generic functions of time).
From eq. (\ref{eq2}) with $h(t)=0$, or equivalently from eq. (\ref{eq3}),
 expressed in nondimensional form,
we obtain
\begin{equation}
\widehat{C}_{vv}(s)=\frac{\langle v^2 \rangle_{\rm eq}}{s+ s \, \widehat{k}(s)}
 \label{eq15}
\end{equation}
From eq.  (\ref{eq4}),  for $\kappa<1$, we have for small $s$ close to $s=0$,
$\widehat{k}(s) \sim 1/s^{1-\kappa}$, and thus
\begin{equation}
\widehat{C}_{vv}(s) \sim \frac{\langle v^2 \rangle_{\rm eq}}{s+s^\kappa}
\sim  \frac{\langle v^2 \rangle_{\rm eq}}{s^\kappa} \, ,   \quad \mbox{for} \;\;\;
s \rightarrow 0
\label{eq16}
\end{equation}
that implies
\begin{equation}
C_{vv}(t) \sim t^{-(1-\kappa)} \, , \qquad t \rightarrow \infty
\label{eq17}
\end{equation}
indicating that, for $\kappa<1$ a superdiffusive diffusional
scaling occurs
\begin{equation}
\langle x^2(t) \rangle \sim t \, \int_0^t C_{vv}(\tau) \, d \tau
\sim t^{1+\kappa}
\label{eq18}
\end{equation}
\begin{figure}
\includegraphics[width=15cm]{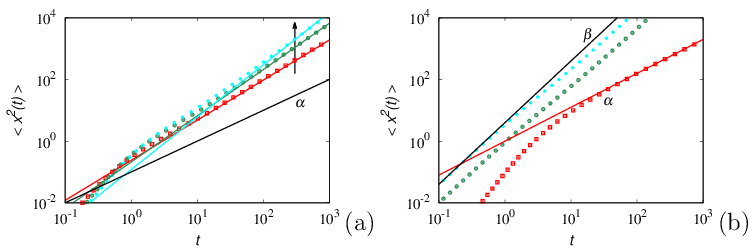}
\caption{$\langle x^2(t) \rangle$ vs $t$, 
  obtained from stochastic simulations
(symbols)
of eqs. (\ref{eq7}) and (\ref{eq14}) for different values
of the exponent $\kappa$, and for different initial conditions
starting from $x(0)=0$. Panel (a) refers to the ergodic case ($\kappa<1$).
The arrow indicates increasing values of $\kappa=0.3,0.5,0.7$.
The solid lines represent eq.  (\ref{eq18}) for the corresponding values of $\kappa$. Line $\alpha$ corresponds to $\langle x^2(t) \rangle \sim t$
in order to mark the difference with respect to the regular diffusive
scaling. The initial conditions are $v(0)=z_n(0)=0$, $n=1,\dots,N$,
but in the ergodic case the initial setting is immaterial.
Panel (b)  refers to $\kappa=2.5$, starting from vanishing
initial condition for the auxiliray variables $z_n(0)=0$, $n=1,\dots,N$
and with $v(0)$ Gaussian random variable with zero mean and variance
$\sigma_0$. Symbols ($\square$) refer to $\sigma_0=0$, symbols ($\circ$)
and ($\bullet$) to $\sigma_0=1,2$, respectively. Line ($\beta$)
corresponds to the ballistic scalings $\langle x^2(t) \rangle \sim t^2$,
while line ($\alpha$) to $\langle x^2(t) \rangle \sim t^{1.1}$.}
 \label{Fig4}
\end{figure}
Conversely, for $\kappa>1$, it is reasonable to argue that, for
generic initial conditions, the scaling of the mean square displacement
should be ballistic, $\langle x^2(t) \rangle \sim t^2$.
This is indeed the case as shown in figure \ref{Fig4}. However,
it should be observed that in the non ergodic case there may exist
initial conditions providing different scalings, at least at intermediate
times, as depicted in figure \ref{Fig4} panel (b) curve ($\alpha$),
starting from vanishing initial conditions for the velocity.

\begin{figure}
\includegraphics[width=15cm]{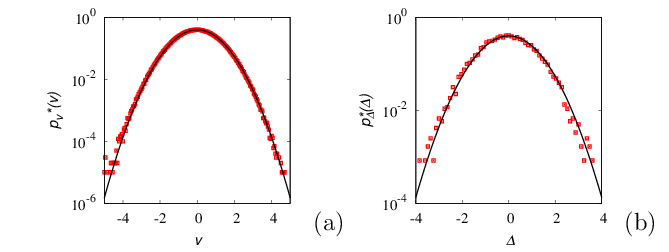}
\caption{Panel (a): equilibrium velocity distribution
$p_v^*(v)$ for $\kappa=0.5$. Panel (b): long-term distribution $p_\Delta^*(\Delta)$ for
the normalized
displacements $\Delta(t)$.  Symbols represent the results
of stochastic simulations, the solid line in both panels
is the normalized Gaussian density with zero mean and unit variance.}
 \label{Fig5}
\end{figure}

In the ergodic case ($\kappa<1$), the statistics of velocity
and normalized  particle displacement $\Delta(t)=(x(t)-x_0)/\sigma_x(t)$,
where $\sigma_x^2(t)= \langle (x(t)-x_0)^2 \rangle$ is 
 Gaussian in the long term,
as depicted in figure \ref{Fig5} for $\kappa=0.5$.

\subsection{Phase transition}
\label{sec3_3}

\begin{figure}
\includegraphics[width=10cm]{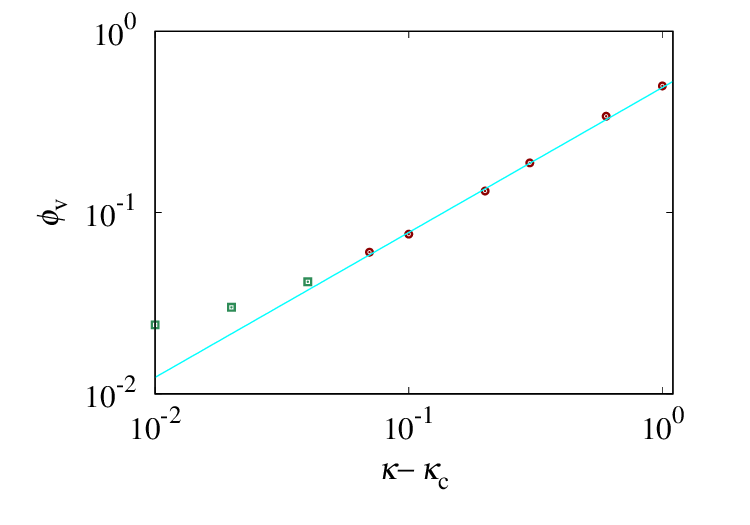}
\caption{Behavior of the order parameter $\phi_v$ 
defined by eq. (\ref{eq19})
as a function of $\kappa-\kappa_c$. Symbols are the results
of moment analysis up to a maximal time order of $10^9$.
Symbols ($\circ$) corresponds to reliable data in this  temporal window.
Symbols ($\square$)  refer to systems for which the
relaxation towards the saturation value of the squared velocity variance
is too slow to ensure an accurate estimate of it at time scales
order of $10^9$. The solid line represent the scaling
$\phi_v \sim (\kappa-\kappa_c)^\zeta$ with $\zeta=0.8$.}
 \label{Fig6}
\end{figure}

The qualitative change of behavior occurring at $\kappa_c=1$
from  ergodic to non-ergodic dynamics is equivalent to a second-order
phase transition.
For this transition we can take 
the normalized difference $\phi_v$ as the order parameter
\begin{equation}
\phi_v = \frac{\langle v^2 \rangle_{\rm eq} - \lim_{t \rightarrow \infty}
\langle v^2(t) \rangle_0}{ \langle v^2 \rangle_{\rm eq} }
\label{eq19}
\end{equation}
i.e. the normalized difference  between the equilibrium value
(in the nondimensional case $\langle v^2 \rangle_{\rm eq}=1$), and the long-term
limit of the squared velocity variance starting from vanishing
initial conditions $v(0)=z_n(0)=0$, $n=1,\dots,N$. The subscript ``$0$''
indicates just this particular initial setting. Of course, other
initial statistical settings, different  from the equilibrium configuration,
can be chosen as well.
Near $\kappa_c$, the order parameter behaves as
\begin{equation}
\phi_v =
\left \{
\begin{array}{ccc}
0 \;\; & \;\;\; & \kappa \leq \kappa_c \\
(\kappa-\kappa_c)^\zeta & & \kappa>\kappa_c
\end{array}
\right .
\label{eq20}
\end{equation}
Figure \ref{Fig6} depicts the behavior of $\phi_v$ as a 
function of $\kappa-\kappa_c$ for $\kappa>\kappa_c$.
Data refer to the solution of the moment equations (\ref{eq12})
starting from $m_{vv}=m_{vz_n}=m_{z_n z_m}=0$, $n,m=1,\dots,N$, (see figure \ref{Fig2})
up to a maximal time order of $10^9$, and are marked differently depending
on their accuracy. Symbols ($\square$)  refer to values of $\kappa$
close to the critical point, for
which the relaxation rate is too slow to ensure an accurate
estimate of the saturation value of the squared velocity variance
at  the timescales considered. Conversely, symbols ($\circ$) refer
to situations where  reliable estimates of the saturation value
were possible.
From the scaling of the latter class of data a value $\zeta=0.8$
of the critical exponent  defined by eq. (\ref{eq20}) is obtained.

\begin{figure} 
\includegraphics[width=10cm]{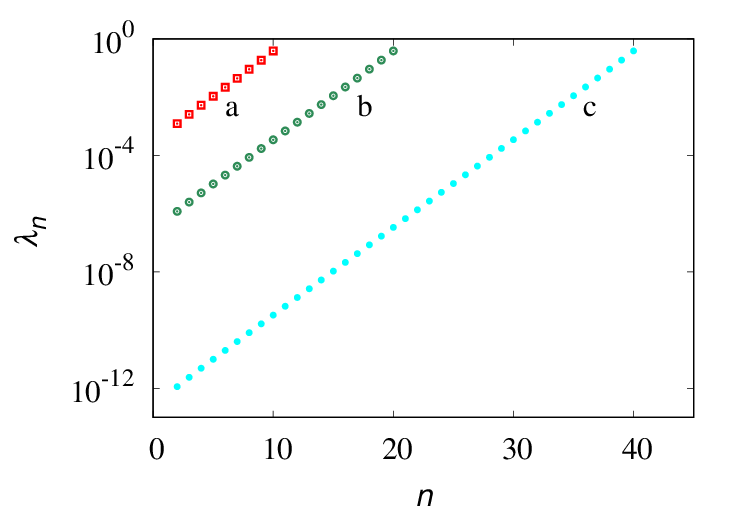}
\caption{Eigenvalues (with reversed sign) $\lambda_n$ vs $n$  associated
with the kernel $k(t)$ defined by the matrix ${\bf A}$ eq. (\ref{eq22})
at $\kappa=0.5$
for different values of the truncation order $N$: (a) $N=10$,
(b) $N=20$, (c) $N=40$.}
 \label{Figs}
\end{figure}

The occurrence of this ergodic/non-ergodic transition can be predicted from
the dynamical properties of the system. Consider eq. (\ref{eq7}) in the
absence of the stochastic forcings, and express it in matrix
form introducing the joint variable ${\bf y}=(v,z_1,\dots,z_N)^T$, where
``$T$ indicates transpose,
\begin{equation}
\frac{d {\bf y}(t)}{d t}= {\bf A}\, {\bf y}(t)
\label{eq21}
\end{equation}
where
\begin{equation}
{\bf A}=
\left (
\begin{array}{ccccc}
-\Gamma & \gamma_1 \mu_1 &  \gamma_2 \mu_2 & \cdots & \gamma_N \mu_N \\
1 & -\mu_1 & 0 & \cdots & 0  \\
1 & 0 & -\mu_2   & \cdots & 0  \\
\cdots & \cdots & \cdots & \cdots & \cdots   \\
1 & 0 & 0 & \cdots & -\mu_N
\end{array}
\right )
\label{eq22}
\end{equation}
Consider the spectral properties of the coefficient matrix ${\bf A}$,
i.e. its right
${\bf A} \, {\bf e}_i= -\lambda_i \, {\bf e}_i$, and
left ${\bf e}_i^* \, {\bf A}= - \lambda_i \, {\bf e}_i^*$,
$i=0,\dots,N$ eigenvectors. The matrix ${\bf A}$ possesses
a vanishing eigenvalue $\lambda_0=0$, while the others
are real and negative, i.e. $\lambda_i>0$, figure \ref{Figs}.

The left eigenvalues constitute the
dual basis, and by definition they are normalized as
\begin{equation}
{\bf e}_i^*({\bf e}_j)= \sum_{k=0}^N e_{i,k}^* e_{j,k} = \delta_{ij}
\label{eq23}
\end{equation}
where $e_{j,k} $ and $e_{i,k}^*$, $k=0,\dots,N$ are the entries
of ${\bf e}_{j}$ and ${\bf e}_i^*$, respectively.
The right and left eigenvectors associated with $\lambda_0=0$ are given by 
\begin{equation}
{\bf e}_0=
\left (
\begin{array}{c}
1 \\
\mu_1^{-1} \\
\mu_2^{-1} \\
\cdots \\
\mu_N^{-1}
\end{array}
\right )
\, ,  \qquad
{\bf e}_0^*
=
 C \, \left ( 1, \gamma_1, \gamma_2, \cdots, \gamma_N \right )
\label{eq24}
\end{equation}
where the constant $C$ is defined by the normalization
condition eq. (\ref{eq23}), i.e. ${\bf e}_0^*({\bf e}_0)=1$,
and thus
\begin{equation}
C= \left ( 1+ \sum_{n=1}^N \frac{\gamma_n}{\mu_n} \right )
\label{eq25}
\end{equation}
Next, consider the evolution of ${\bf y}(t)$, solution
of eq. (\ref{eq21}) starting from a generic
initial condition ${\bf y}(0)={\bf y}_0$,
\begin{equation}
{\bf y}(t)= e^{{\bf A} \, t} \, {\bf y}_0 = \sum_{n=0}^N
{\bf e}_n^*({\bf y}_0) \, e^{-\lambda_n \, t} {\bf e}_n
\label{eq26}
\end{equation}
In the long-term limit, ${\bf y}(t)$ converges to a point
lying in the central eigenmanifold spanned by ${\bf e}_0$,
\begin{equation}
\lim_{t \rightarrow \infty} {\bf y}(t)= {\bf e}_0^*({\bf y}_0) \, {\bf e}_0
\label{eq27}
\end{equation}
depending on the initial condition.
From eqs. (\ref{eq24})-(\ref{eq25}),  setting ${\bf  y}_0=(v_0,z_{1,0},\dots,
z_{N,0})$, we have
\begin{equation}
{\bf e}_0^*({\bf y}_0) = \frac{v_0 + \sum_{n=1}^N \gamma_n \, z_{n,0}}{1+ \Phi_N}
\label{eq28}	
\end{equation}
where
\begin{equation}
\Phi_N = 1+ \sum_{n=1}^N \frac{\gamma_n}{\mu_n} = 1+ \sum_{n=1}^N \frac{1}{2^{n (\kappa -1)}}
\label{eq29}
\end{equation}
Two situations occur in the limit for $N \rightarrow \infty$:
(i) if $\kappa<1$, $\lim_{N \rightarrow \infty} \Phi_N = \infty$.
Consequently, ${\bf e}_0^*({\bf y}_0)={\bf 0}$, and the
long-term limit of the solution eq. (\ref{eq26}), is independent of the initial
condition ${\bf y}_0$. This is the case providing the occurrence
of an ergodic behavior; (ii) if $\kappa>1$, $\lim_{N \rightarrow \infty}
\Phi_N= \Phi_\infty < \infty$, and the long-term limit eq. (\ref{eq27})
does depend on the initial conditions, corresponding to a non ergodic
behavior.

It can be observed that the condition on $\Phi_N$ in the
limit for $N \rightarrow \infty$ corresponds to the
integrability condition  for the inertial kernel $k(t)$.
Indicating with $k_N(t)$ the truncated approximation of $k(t)$ up to
$N$ modes,  we have
\begin{equation}
\Phi_N= 1+ \int_0^\infty k_N(t) \, d t
\label{eq30}
\end{equation}

\begin{figure}
\includegraphics[width=10cm]{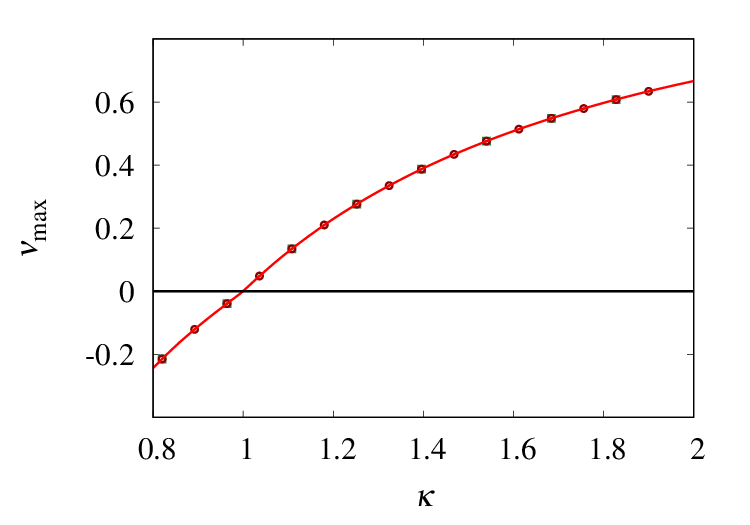}
\caption{Maximal real part $\nu_{\rm max}$  of the eigenvalues
associated with the coefficient matrix of the linear reduced system
eq. (\ref{eq32}) as the exponent $\kappa$ varies.}
 \label{Fig7}
\end{figure}

There is also another way to detect the ergodicity transition,
adopting a quasi steady-state approximation of the dynamics.
Consider the dynamics of the second-order moments eq. (\ref{eq12}),
and perform a quasi steady-state approximation for $m_{vv}$ and $m_{z_m z_p}$,
corresponding to  the assumption that $d m_{vv}/dt= d m_{z_m z_p}/d t=0$.
This leads to the following algebraic conditions,
\begin{eqnarray}
m_{vv} & = & \frac{1}{\Gamma} \sum_{n=1}^N \gamma_n \, \mu_n \, m_{v z_n}
+ \frac{1}{\Gamma} \sum_{n=1}^N d_n^2 \\
m_{z_m z_p}& =  & \frac{m_{v z_m} + m_{v z_p}+ 2 \, c_n^2 \delta_{mp}}{\mu_m + \mu_p}
\label{eq31}
\end{eqnarray}
Substituting these expression into the dynamics for the mixed moments
$m_{v z_n}(t)$ (second set of equations in eq. (\ref{eq12})), and
enforcing the definition of the coefficients $d_n$, $c_n$ eq. (\ref{eq8}),
one
ends up with the reduced dynamics
\begin{eqnarray}
\frac{d m_{v z_n}}{d t} & = & - (\Gamma + \mu_n) \, m_{v z_n}+
\sum_{m=1}^N \frac{\gamma_m \, \mu_m}{\mu_n+ \mu_m} m_{v z_n}
+ \sum_{m=1}^N \frac{\gamma_m \, \mu_m}{\mu_n+ \mu_m} m_{v z_m} \nonumber \\
& + & \frac{1}{\Gamma} \sum_{m=1}^N \gamma_m \, \mu_m \, m_{v z_m}
\label{eq32}
\end{eqnarray}
Let ${\bf B}$ the constant $N \times N$ coefficient matrix
associated with the reduced model eq. (\ref{eq32}) and let $\nu_{\rm max}$
be the maximal real part of its eigenvalues.
The graph of $\nu_{\rm max}$ vs $\kappa$ is depicted in figure 
\ref{Fig7}.
For $\kappa<1$,  all the eigenvalues have negative real
part. At $\kappa=\kappa_c=1$ the reduced system loses stability and
$\nu_{\rm max}$ becomes positive for $\kappa>1$.
Although there is no specific dynamic motivation underlying
the quasi steady-state assumption, the physical
interpretation of this result is  rather clear.
Since for  $\kappa<1$ all the eigenvalues of ${\bf B}$ are negative,
the mixed moments $m_{v z_n}(t)$ are vanishing in the
long term. The vanishing of these mixed moments is the
condition (detailed Langevin condition) ensuring the validity of Kubo fluctuation-dissipation relations (see Appendix A), and in turn 
$\langle v^2(t) \rangle$  would
converge to the equilibrium value $\langle v^2 \rangle_{\rm eq}=1$,
independently of the initial conditions (ergodicity).
Conversely, the instability characterizing the reduced
system eq. (\ref{eq32}) for the mixed moments with $\kappa>1$
indicates that in this case $m_{v z_n}(t)$ cannot be
vanishing in the long term, and consequently the dynamics
of the system  behaves in a way not consistent with
the fluctuation-dissipation predictions. The latter analysis
of the transition occurring at $\kappa=\kappa_c$ recasts the problem in the form of a bifurcational
issue associated with the loss of stability of an equilibrium
point  for the second-order reduced moment dynamics.

\subsection{Presence of attractive potentials}
\label{sec3_4}

Another interesting situation involves the presence of attractive potentials
$U(x)$ such that $\int_{-\infty}^\infty e^{-U(x)} \, d x < \infty$,
ensuring the existence of a steady state. Observe that in the
nondimensional formulation $k_B \, T =1$.
In this case, the first  eq. (\ref{eq7})  for $d v(t)/dt$ should be
substituted by
\begin{equation}
\frac{d v(t)}{d t}= - \Gamma \, v(t) + \sum_{n=1}^N \gamma_n \, \mu_n \, z_n(t)
- \frac{d U(x(t))}{d x}+ \sqrt{2} \, \sum_{n01}^N d_n \, \xi_n(t)
\label{eq33}
\end{equation}
coupled with the kinematic equation eq. (\ref{eq14}).
We consider two prototypical attractive potentials: (i) the
harmonic potential
\begin{equation}
U(x)= \frac{x^2}{2}
\label{eq34}
\end{equation}
and the double-well potential
\begin{equation}
U(x)= \frac{x^4}{4}- \frac{x^2}{2}
\label{eq35}
\end{equation}
In both cases, the long term steady-state solution for the marginal velocity
and position densities are the Maxwellian and the Boltzmannian respectively.
This is shown in figure \ref{Fig8}.
This result is independent of the initial conditions (see further data
reported in Appendix B)  and is valid for any value of $\kappa$.
\begin{figure}
\includegraphics[width=15cm]{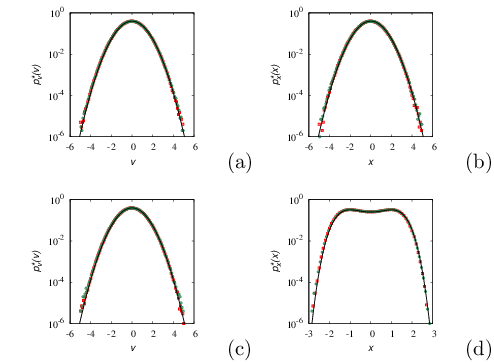}
\caption{Equilibrium velocity $p_v^*(v)$ (panels (a) and (c)),
and position $p_x^*(x)$ (panels (b) and (d)) marginal densities 
 in the presence of attractive potentials.
Symbols represent the results of stochastic simulations: ($\square$)
correspond to $\kappa=0.5$, ($\circ$) to $\kappa=2.5$.
Panels (a) and (b) refer to the harmonic potential, panels
(c) and (d) to the double-well potential.
Solid lines in panels (a), (b) and (c) correspond to
the Gaussian distribution with zero mean and unit variance.
Solid line in panel (d) is the Boltzmann distribution 
$p_x^*(x)=A e^{-U(x)}$, where $A$ is the normalization constant and
$U(x)=x^4/4-x^2/2$ is the double-well potential.}
 \label{Fig8}
\end{figure}
Indeed, this result is valid in the  presence of any dissipationless
inertial kernel. To prove this, we can consider the simplest possible
case, namely an exponential kernel $k(t)=\gamma e^{-\mu \, t}$
defined by a single mode, in the presence of a harmonic potential.
The modal equations for this setting read
\begin{eqnarray}
\frac{d x(t)}{d t} & = & v(t)\nonumber \\
\frac{d v(t)}{d t} & =  & - \gamma \, v(t) + \gamma \, \mu \, z(t) -x(t)
+\sqrt{2} \, d \, \xi(t) \label{eq36} \\
\frac{d z(t)}{d t} & = &- \mu \, z(t) + v(t) + \sqrt{2}\, c \, \xi(t)
\nonumber
\end{eqnarray}
where $d=\sqrt{\gamma}$ and $c=-1/\sqrt{\gamma}$
 as addressed in Appendix B.
If we consider the dynamics of the second-order moments we obtain
at steady state (steady-state moments are indicated with an asterix),
\begin{equation}
m_{vv}^*=m_{xx}^* = 1 \,, \quad m_{zz}^*=\frac{1}{\gamma \, \mu} \, ,
\quad
m_{xv}^*=m_{xz}^* =m_{vz}^* = 0
\label{eq37}
\end{equation}
corresponding to the equilibrium solution. This result can be
straightforwardly generalized to the presence of an arbitrary number
of modes. This means that the presence of an attractive potential
(the theoretical analysis is limited to the harmonic case)
provides a form of thermalization even in the absence of
dissipative effects deriving from fluid-particle interactions.

The explanation of this phenomenon is purely of dynamic nature.
Consider  again the one-mode model eq. (\ref{eq36}) which is governed
by a linear dynamic for ${\bf y}=(x,v,z)$ controlled by
the coefficient matrix ${\bf A}$
\begin{equation}
{\bf A}=
\left (
\begin{array}{ccc}
\; 0  \; & \; 1 \; & \; 0 \; \\
-1 & - \gamma & \, \gamma \; \mu \\
0 & 1 & \; -\mu
\end{array}
\right )
\label{eq38}
\end{equation}
From the analysis of the characteristic polynomial associated
with the matrix ${\bf A}$, it follows that all the eigenvalues
admit negative real part for $\gamma, \, \mu>0$. Therefore, the presence of
a harmonic potential introduces dynamically a stabilizing 
behavior enabling the achievement of a thermalizing action
of thermal fluctuations even in the absence of dissipative
fluid inertial interactions.

\subsection{Presence of steady forces}
\label{sec3_5}

As addressed in the previous paragraph there are some features of the class
of inertial kernel defined by eqs. (\ref{eq3})-(\ref{eq4}) that
make them acquire some irreversible behavior leading to
thermalization and to an equilibrium solution independently of
the initial conditions in the presence of attractive potentials.

This stems essentially from the spectral structure of the
resulting GLE defined by the matrix ${\bf A}$  eq. (\ref{eq22}):
it possesses a vanishing eigenvalue while the remaining ones are strictly
negative (see figure \ref{Figs}).
The presence of a harmonic potential destroys the central
eigenspace, turning the central eigenvalue from vanishing to
negative.

In this regard, there is a profound difference with respect to
the kernels arising from linear heat-bath models (Zwanzig heat-baths) \cite{zw1,zw2,hb1} involving
a finite but  arbitrarily large numbers of bath particles, for which
all the eigenvalues of the coefficient matrix are purely
imaginary, and the superposition of an external potential
acting on the tagged particle does not  change the qualitative (spectral)
properties of the dynamics.

However, there is a clear an unambiguous indicator
of dissipation that marks the difference between truly dissipative
(strong thermalizing) and apparently dissipative (weak thermalizing)
systems: the  long-term velocity dynamics in a presence
of a bounded force and specifically a constant one.
For this reason, we consider this case, for which the
first  eq. (\ref{eq7})  for $d v(t)/dt$  becomes
\begin{equation}
\frac{d v(t)}{d t}= - \Gamma \, v(t) + \sum_{n=1}^N \gamma_n \, \mu_n \, z_n(t)
+ F+ \sqrt{2} \, \sum_{n=1}^N d_n \, \xi_n(t)
\label{eq39}
\end{equation}
where $F>0$ is constant.

\begin{figure}
\includegraphics[width=15cm]{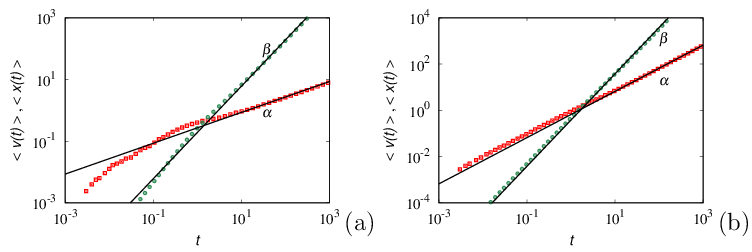}
\caption{Mean values for velocity $\langle v(t) \rangle$ (symbols ($\circ$) and lines $\alpha$)
and position $\langle x(t) \rangle$ (symbols ($\square$) and lines $\beta$)
 vs time $t$ in the presence of a constant force $F=1$.  Panel (a) refes to $\kappa=0.5$,
panel (b)to $\kappa=2.5$). Symbols are the results of stochastic
simulations. Line $\alpha$ in panel (a) corresponds to
$\langle v(t) \rangle \sim t^\kappa$, line $\beta$ to
$\langle x(t) \rangle \sim t^{1+\kappa}$.
Line $\alpha$ in panel (b) corresponds to
$\langle v(t) \rangle \sim t$, line $\beta$ to
$\langle x(t) \rangle \sim t^2$.
}
 \label{Fig9}
\end{figure}

In the dissipative case, the particle velocity relaxes towards
a constant value, the settling velocity $v_\infty=F/\eta_\infty$,
where $\eta_\infty=\int_0^\infty h(t) \, d t$ is the integral
of the dissipative memory kernel $h(t)$ introduced in eq. (\ref{eq2}).

The saturation towards a constant settling velocity
cannot happen in the absence of dissipation, and
the particle velocity grows unboundenly in time.

This can be easely seen from eq. (\ref{eq39}), considering
the first-order moment, i.e. the expected value
for the velocity.
In the Laplace domain we have for small $s$ (long times)
\begin{equation}
\langle \widehat{v}(s) \rangle = \frac{F}{s \, (s+ s \, \widehat{k}(s))}
\sim \frac{F}{s  \, (s + s^\kappa)}
\label{eq40}
\end{equation}
Also in this case a different scaling occurs  below and above
the critical value $\kappa_c=1$.
For $\kappa<1$, the term $s^\kappa$ within the brackets at
the denominator of eq. (\ref{eq40})  is the leading term
with respect to $s$ for small values of $s$, thus
$\langle \widehat{v}(s) \rangle \sim F/s^{1+\kappa}$,
leading to  $\langle v(t) \rangle  \sim  F \, t^\kappa$,
 $\langle x(t) \rangle  \sim  F \, t^{1+\kappa}$.
For $\kappa>1$, the opposite occurs, 
$\langle \widehat{v}(s) \rangle \sim F/s^2$,
and as a consequence $\langle v(t) \rangle  \sim  F \, t$,
 $\langle x(t) \rangle  \sim  F \, t^2$. This phenomenon
is depicted in figure \ref{Fig9} with the aid of data deriving
from stochastic simulations of eq. (\ref{eq39}) using
an ensemble of $10^5$ realizations.

\section{Weak thermalization and strong dissipation: a discussion on
irreversibility}
\label{sec4}

The results developes in the previous Section require a deeper analysis
as regards the connection between irreversibility, thermalization and dissipation,  as some form of thermalization and irreversibility is 
exhibited by non-dissipative system (associated with
power-law inertial kernels with $\kappa<1$).
For the sake of clarity, it is convenient to introduce some definitions
taking the GLE associated with particle hydromechanics as a working
example. We refer  to  isothermal conditions.

A GLE is {\em weakly thermalizing} if, in the  absence
of external perturbations (external forces and potentials) other
than the equilibrium thermal force, the velocity attains
a unique equilibrium distribution independently of the initial
state of the system.

This implies that $\langle v^2(t) \rangle$ would converge
for long times to the equilibrium value $\langle v^2 \rangle_{\rm eq}$,
independently of the initial preparation of the system, i.e.
of the   values for $v(t)$ and $z_n(t)$ at time $t=0$.

A GLE is {\em canonically robust} if, in the presence of attractive
potentials $U(x)$ superimposed to
the  equilibrium thermal  fluctuations, there exists a unique
equilibrium distribution, independently of the initial conditions,
coinciding with the Maxwell-Boltzmann distribution.

This diction stems  from the property that the GLE provides
equilibrium statistics analogous to those predicted by the
equilibrium ensemble theory (canonical ensemble).

A GLE is {\em strongly thermalizing} if it
is weakly thermalizing and canonically consistent and moreover, for any bounded force $F(x)$
(such that there exists a constant $K>0$ and $|F(x)| < K$) superimposed
to the equilibrium thermal fluctuations, the velocity attains in the long-term
limit a bounded value, independently of the initial conditions.
More specifically, if the force applied is constant, the velocity
slould converge towards a constant limit value.

In the light of these definitions the behavior displayed by  GLEs
characterized by dissipationless inertial kernel with long-term
power-law scaling can be summarized as follows:
\begin{itemize}
\item for $\kappa<1$ the GLE is both weakly thermalizing
and canonically robust;
\item for $\kappa>1$ the GLE is canonically robust but  the
property of weak thermalization is lost;
\item independently of the value of $\kappa$ the GLE is not strongly
thermalizing
\end{itemize}
Therefore, in the absence of dissipation it may happen than a GLE
may  display thermalization properties analogous to those
of dissipative systems, i.e. some form of ``irreversibility''.
Nevertheless, the most strict property characterizing dissipative
dynamics is their behavior in the presence  of bounded forces.
And this  represents the most cogent test for a correct thermodynamic
behavior. To this end, we have shown that dissipationless
GLE cannot provide this type of limit behavior.
The simplest physical model is the sedimentation of a solid particle
in a fluid
(with density greater than the density of the fluid medium) under
the action of gravity. In the experiments a final settling
velocity is achieved, supporting the claim that fluid-particle
interactions are  strictly dissipative. 

It is important to
observe, that using some mathematical technicalities, 
it would be possible to justify the occurrence of
a thermalizing behavior in dissipationless systems  
even in situations where thermalization cannot be physically
achieved.
This  is associated with the existence of singular limits,
and with the way these limits are performed.

Let us address this topic by means of a simple example
consistent with the class of problems analyzed in this
article.
Consider a GLE characterized by an exponential
kernel $k(t)=\gamma e^{-\mu \, t}$, and by a
small friction factor $\epsilon>0$. With reference to eq. (\ref{eq2})
this implies that the dissipative kernel $h(t)$ is impulsive, $h(t)=\varepsilon
\, \delta(t)$.
In this case, the modal representation consistent with Kubo
fluctuation-dissipation theory reads (see Appendix B)
\begin{eqnarray}
\frac{d v(t)}{d t} & = & - (\varepsilon+\gamma) \, v(t) + \gamma \, \mu \, z(t)
+ \sqrt{2 \, \varepsilon}  \, \xi^\prime(t) + \sqrt{2} \, d \, \xi(t)
\nonumber \\
\frac{d z(t)}{d t} & = & - \mu \, z(t) + v(t) + \sqrt{2} \, c \, \xi(t)
\label{eqa1}
\end{eqnarray}
where $d=\sqrt{\gamma}$, $c=-1/\sqrt{\gamma}$ and $\xi(t)$, $\xi^\prime(t)$
are distributional derivatives of two independent Wiener processes.
For $\varepsilon>0$, the system is dissipative and thus, for
any initial conditions (expressed via the statistical
properties of $v(0)=v_0$ and $z(0)=z_0)$, one obtains
\begin{equation}
\lim_{t \rightarrow \infty} \langle v^2(t) \rangle =1
\label{eqa2}
\end{equation}
Consequently,
\begin{equation}
\lim_{\varepsilon \rightarrow 0} \lim_{t \rightarrow \infty} \langle v^2(t) \rangle =1
\label{eqa3}
\end{equation}
Alternatively, one can exchange the order of execution of the limits
in eq. (\ref{eqa3}), taking firstly the limit for $\varepsilon \rightarrow 0$,
and subsequently the limit for $t \rightarrow \infty$.
Doing so, one obtains
\begin{equation}
 \lim_{t \rightarrow \infty} \lim_{\varepsilon \rightarrow 0} \langle v^2(t) \rangle = f[p_0(v_0,z_0)]
\label{eqa4}
\end{equation}
where $f[p_0(v_0,z_0)]$ is a functional of
the statistical properties of the initial preparation of the system
defined by the initial joint probability density function $p_0(v_0,z_0)$.
Specifically,
\begin{equation}
f[p_0(v_0,z_0)]= 1+ \mu \frac{2 \, \gamma \, \mu \, m_{vz}(0) +
\mu \, m_{vv}(0) + \gamma^2 \, \mu \, m_{zz}(0)}{(\gamma+\mu)^2}
\label{eqa5}
\end{equation}
where
\begin{equation}
m_{\alpha \beta}(0)=\int \int \alpha \, \beta p_0(v_0,z_0) \, d v_0 \, d z_0
\,, \qquad \alpha,\beta=v_0,z_0
\label{eqa6}
\end{equation}
are the second-order moments of the initial distribution.
Eq. (\ref{eqa5}) stems from elementary moment analysis.
We may therefore conclude that
\begin{equation}
\lim_{t \rightarrow \infty} \lim_{\varepsilon \rightarrow 0} \langle v^2(t) \rangle 
\neq \lim_{\varepsilon \rightarrow 0} \lim_{t \rightarrow \infty} \langle v^2(t) \rangle 
\label{eqa7}
\end{equation}

The qualitative properties displayed by the singular
limit in eq. (\ref{eqa7}) are altogether analogous to other
singular limits occurring in the theory of GLEs and dissipation.
Specifically, they occur in the definition of the memory kernel
associated with Hamiltonian heat bath models \cite{hb1}, where the limit
for time tending to infinity and for the number of heat-bath
particles tending to infinity may not coincide.
In the latter case, and in the case of eq. (\ref{eqa7}),
it is the physical reasoning that may resolve the apparent
ambiguity of the limit,  providing the correct interpretation
of it.

\section{Concluding remarks}
\label{sec5}
In order to complete the analysis of ergodicity breaking in GLEs initiated
with the work \cite{gpp2}, we have explored in this article
the case of dissipationless  kernels. In the hydromechanic framework
this correspond to the presence of purely fluid inertial
interactions. The case of  inertial kernels with
long-term power-law tails is particularly interesting as
a phase transition occurs at the critical value  $\kappa=1$ of the
exponent characterizing  the long-term behavior.

For $\kappa<1$ the system  displays weak thermalization
and is ergodic both in the presence of purely thermal fluctuations
and in the case attractive potentials are superimposed to it.
Weak thermalization is lost for $\kappa>1$ together
with the breakdown of ergodicity.
For $\kappa<1$ a superdiffusive scaling of the mean square displacement
is observed with an exponent equal $1+\kappa$.

While the results for $\kappa<1$ seems to suggest that, even in
the absence of dissipation, some form of irreversibility may arise,
the critical test for the latter claim involves the  dynamic
properties of momentum transfer
in the presence of constant or bounded potentials.
And enforcing this test (for a constant force) in the absence
of a real dissipative fluid-particle interaction mechanism,
the velocity diverges to infinity, and by no mean it can reach
a constant and bounded settling velocity as observed e.g. in
sedimentation experiments. In other words, even if for
$\kappa<1$ the dissipationless GLE  is weak thermalizing,
this is not true in the strong sense, and consistently
it cannot describe the thermodynamic and energetic properties
of an irreversible memory dynamics.\\

\vspace{0.3cm}
\noindent
\appendix*{\bf Appendix A}

Consider the hydromechanic model for fluid-particle
interactions characterized by a constant Stokesian
friction factor $\varepsilon$ and by a fluid inertial
kernel $k(t)$ of the form defined by eq.  (\ref{eq5}).
Its modal expansion reads
\begin{eqnarray}
\frac{d v(t)}{d t} & = & - (\varepsilon+\Gamma) \, v(t)+\sum_{n=1}^N \gamma_n \, \mu_n \, z_n(t)
+ \sqrt{2} \, a \, \xi^\prime(t) + \sqrt{2} \, \sum_{n=1}^N d_n \, \xi_n(t) \nonumber \\
\frac{d z_n(t)}{d t}& = &- \mu_n \, z_n(t) + v(t) + \sqrt{2} \, c_n \, \xi_n(t)
\label{eqb1}
\end{eqnarray}
where $\xi^\prime(t)$, $\xi_n(t)$, $n=1,\dots,N$ are distributional
derivatives of independent Wiener processes, and the coefficient $a$, $d_n$, $c_n$ should
be determined  by fluctuation-dissipation theory.

For $\varepsilon>0$ the system is ergodic: there exists a unique stationary density
of the joint process $(v(t),z_1(t),\dots,z_n(t))$ for any
values of the coefficients $a$, $d_n$, $c_n$, $n=1,\dots,N$.
These coefficients can be determined by enforcing the first Kubo fluctuation-dissipation
relation, namely that the velocity autocorrelation function $C_{vv}(t)$ should
be the solution of the equation
\begin{equation}
\frac{d C_{vv}(t)}{d t}= - (\varepsilon+\Gamma) \, C_{vv}(t)+ \sum_{n=1}^N \gamma_n \, \mu_n \, e^{-\mu_n \, t}
* C_{vv}(t)
\label{eqb2}
\end{equation}
where ``$*$'' indicates convolution, and $C_{vv}(0)=\langle v^2 \rangle_{\rm eq}=1$.
Consider the dynamics of the second-order moments $m_{vv}(t)=\langle v^2(t)\rangle$,
 $m_{vz_n}(t)=\langle v(t) z_n(t) \rangle$,  $m_{z_n z_m}(t)=\langle z_n(t)  z_m(t) \rangle$
associated with eq. (\ref{eqb1}), and the corresponding velocity autocorrelation function.
It is easy to see that $C_{vv}(t)$ satisfies eq. (\ref{eqb2}) if and only if
$m_{vv}^*=1$, and
the second mixed moments at equilibrium (indicated by an asterix) satisfy the conditions
\begin{equation}
m_{v z_n}^*=0 \, , \qquad n=1,\dots, N
\label{eqb3}
\end{equation}
that represent the detailed Langevin conditions in the modal expansion.
Enforcing eq. (\ref{eqb3}) for the moment values at equilibrium, the expression
for the coefficients $a$, $d_n$, $c_n$, $n=1,\dots,N$ can be obtained, namely
\begin{equation}
a=\sqrt{\varepsilon} \, , \quad d_n=\sqrt{\gamma_n} \, , \quad c_n=-\frac{1}{\sqrt{\gamma_n}} \, ,
\quad n=1,\dots, N
\label{eqb4}
\end{equation}
The dissipationless case  consistent with the Kubo fluctuation-dissipation theory
is obtained from   eqs. (\ref{eqb1}), (\ref{eqb4}) for $\varepsilon=0$,  corresponding
to eqs. (\ref{eq7})-(\ref{eq8}) in the article.\\

\vspace{0.3cm}
\noindent
\appendix*{\bf Appendix B}

In this appendix we report some additional simulation results supporting the
analysis developed in the article.

Consider the thermalization dynamics i.e. eq. (\ref{eq7}) starting from different initial conditions and using an ensemble of $10^5$ realizations.

\begin{figure}
\includegraphics[width=15cm]{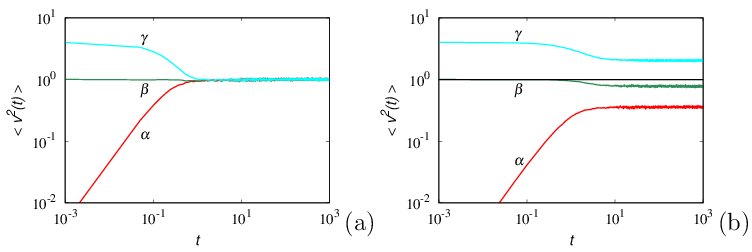}
\caption{$\langle v^2(t) \rangle$ vs $t$ for initial conditions $z_n(0)=0$ $n=1,\dots,N$,
and  $v(0)$ normally distributed with variance $\sigma_0$. Panel (a)
refers to $\kappa=0.5$, panel (b) to $\kappa=2.5$.
Lines $\alpha$, $\beta$, $\gamma$ correspond to $\sigma_0=0,1,\sqrt{2}$, respectively.
}
 \label{Fig10}
\end{figure}

Figure \ref{Fig10} depicts the behavior of $\langle v^2(t) \rangle$, in the case $z_n(0)=0$ $n=1,\dots,N$,
and $v(0)$ is a Gaussian random variable with zero mean and variance $\sigma_0$ for GLEs within and outside
the ergodicity region ($\kappa<1$).

\begin{figure}
\includegraphics[width=15cm]{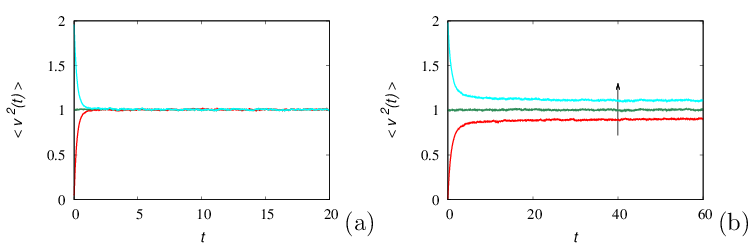}
\caption{$\langle v^2(t) \rangle$ vs $t$ for initial conditions where
 $z_n(0)=0$ $n=1,\dots,N$ are normally distributed with zero
mean and variances equal to their equilibrium varianced,
and  $v(0)$ normally distributed with variance $\sigma_0$. Panel (a)
refers to $\kappa=0.5$, panel (b) to $\kappa=1.5$.
Lines $\alpha$, $\beta$, $\gamma$ correspond to $\sigma_0=0,1,\sqrt{2}$, respectively.}
 \label{Fig11}
\end{figure}

Another set of initial conditions is depicted in figure \ref{Fig11}.
In this case $v(0)$ is still Gaussian with variance $\sigma_0$,
while $z_n(0)$ $n=1,\dots,N$ are Gaussian random variables
sampled from their equilibrium distribution: a Gaussian
distribution with zero mean and variance $1/\sqrt{\gamma_n \, \mu_n}$, 
$n=1,\dots,N$.
In the non-ergodic case ($\kappa=1.5$, panel (b) of figure \ref{Fig11})
if $\sigma_0=1$, i.e. if the initial condition coincides with
the thermodynamic equilibrium conditions the system remains
in equilibrium. In all the other cases $\sigma_0 \neq 1$,
the limiting value of $\langle v^2(t) \rangle$ is different
from $\langle v^2 \rangle_{\rm eq}=1$.

\begin{figure}
\includegraphics[width=15cm]{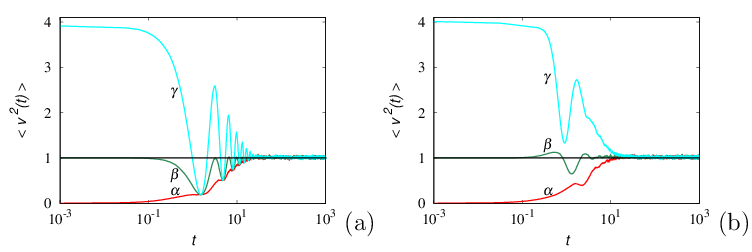}
\caption{$\langle v^2(t) \rangle$ vs $t$ in the presence of
attractive potentials, $\kappa=2.5$, for initial conditions $z_n(0)=0$ $n=1,\dots,N$, 
and  $v(0)$ normally distributed with variance $\sigma_0$.
Panel (a)
refers to the harmonic potential, panel (b) to the double-well potential.
Lines $\alpha$, $\beta$, $\gamma$ correspond to $\sigma_0=0,1,2$, respectively.
}
 \label{Fig12}
\end{figure}

Finally, let us consider  the thermalization in the presence of
the attractive potentials (harmonic and double-well potential
consider in the main text) for values of $\kappa>1$.
In this case, the presence of attractive potentials acts as a ``thermalizator''
in the meaning that independently of the initial conditions
the squared velocity variance reaches its equilibrium value
$\langle v^2 \rangle_{\rm eq}=1$.
This phenomenon is depicted in figure \ref{Fig12}.


\vspace{0.3cm}
\begin{thebibliography}{50}

\bibitem{kubo1}  R. Kubo,   Rep.  Prog.  Phys. {\bf 29}, 255 (1966).
\bibitem{kubo2} R. Kubo, M. Toda and N. Hashitsune, {\em Statistical Physics II Nonequilibrium
Statistical Mechanics} (Springer-Verlag, Berlin, 1991).
\bibitem{gle1}  H. Mori,  
Prog. Theor. Phys. {\bf 33}, 423 (1965). 
\bibitem{gle2} R. M. Hill and L. A. Dissado, J. Phys. C {\bf 18},
3829 (1985).
\bibitem{gle3} L. Stella, C. D.Lorenz and L. Kantorovich,
Phys. Rev. B {\bf 89}, 134303 (2014).
\bibitem{gle4} T. Schilling, 
Phys. Rep. {\bf 972}, 1 (2022).


\bibitem{gen}  O. Darrigol, Eur. Phys. J. H {\bf 48} 10 (2023).
\bibitem{br1} J.-D. Bao and Y.-Z. Zhuo,
Phys. Rev. Lett. {\bf 91}, 138104 (2003).
\bibitem{br2} J.-D. Bao, P. H\"anggi and Y.-Z. Zhuo,
Phys. Rev. E {\bf 72}, 061107 (2005).
\bibitem{br3} P. Siegle, I. Goychuk and  P. H\"anggi, EPL {\bf 93}, 20002 (2011).

\bibitem{br4} F. Ishikawa and S. Todo, Phys. Rev. E {\bf 98}, 062140 (2018).
\bibitem{plyukhin}  A. V. Plyukhin, Phys.  Rev. E {\bf 83}, 062102 (2011).
\bibitem{ct1} . Bel and E. Barkai, Phys. Rev. Lett.
{\bf 94}, 240602 (2005).
\bibitem{ct2} R. Metzler, J.-H. Jeon, A. G. Cherstvy
 and E. Barkai, Phys. Chem. Chem. Phys. 
{\bf 16}, 24128 (2014).
\bibitem{ct3} Y. Liang, W. Wang, R. Metzler and A. G.
Cherstvy, 
Phys. Rev. E {\bf 108}, 034113 (2023).
\bibitem{gpp1} M. Giona, G. Procopio and C. Pezzotti,
Phys. Rev. E {\bf 111}, 034105 (2025).
\bibitem{gpp2} G. Procopio, C. Pezzotti and M. Giona,
 Phys. Rev. E {\bf 111} 034106 (2025).
\bibitem{hb1} M. Giona, G. Procopio amd C. Pezzotti,
http://dx.doi.org/10.2139/ssrn.5039729 (2024).
\bibitem{rheol1} C. W. Macosko,  {\em Rheology - Principles, Measurements, and Applications}
(Wiley-VCH, New York, 1994).
\bibitem{rheol2} J. D. Ferry, {\em Viscoelastic Properties of Polymers} (J. Wiley \& Sons, New York, 1970).
\bibitem{landau}  L. D.  Landau and E. M. Lifshitz, {\em Fluid Mechanics}; (Pergamon Press, Oxford, 1993).

\bibitem{pofgiona} M. Giona,  G. Procopio and C. Pezzotti. 
Phys.  Fluids {\bf 37} 022042 (2025).

\bibitem{raizen1} R. Huang, I. Chavez, K. M. Taute, B. Lukic, S. Jeney,
M. G.  Raizen and  E. L.  Florin,   
 Nature Phys. {\bf 7},
 576 (2011).

\bibitem{raizen2} J. Mo  and M. G. Raizen,
Annu. Rev. of Fluid Mech. 51 403 (2019).
\bibitem{peppevisco} G. Procopio and M. Giona,
 Fluids {\bf 8} 84 (2023).

\bibitem{mason1} T. G. Mason, K. Ganesan, J. H. van Zanten, D. Wirtz 
and S. C.  Kuo,
 Phys. Rev. Lett. {\bf 79}, 32825 (1997).

\bibitem{mason2}  T. M. Squires and T. G. Mason,  
Annu. Rev. Fluid Mech. {\bf 42} 413 (2010).

\bibitem{zw1} R. Zwanzig,  J. Stat. Phys.  {\bf 9} 215 (1973).
\bibitem{zw2} R. Zwanzig, {\em Nonequilibrium Statistical Mechanics} (Oxford Univ. Press, Oxford, 2001).

\end{thebibliography}
\end{document}